\documentclass[american,superscriptaddress,aps,reprint,prl]{revtex4-2}
\usepackage[T1]{fontenc}
\usepackage[utf8]{inputenc}
\usepackage{bm}
\usepackage{amsmath}
\usepackage{amssymb}
\usepackage{comment}
\usepackage{graphicx}
\usepackage{etoolbox}
\usepackage{xcolor}
\usepackage{babel}
\usepackage{newtxtext}

\apptocmd{\thebibliography}{\sloppy}{}{}
\begin{document}

\author{Jonathan Bauermann}
\email{jbauermann@fas.harvard.edu}
\affiliation{Department of Physics, Harvard University, Cambridge, MA 02138, United States of America}

\author{Roberto Benzi}
\email{roberto.benzi@gmail.com}
\affiliation{Sino-Europe Complex Science Center, School of Mathematics Northwestern University of China, Shanxi, Taiyuan 030051, China}
\affiliation{Department of Physics and INFN, University of Rome Tor Vergata, I-00133 Rome, Italy}

\author{David R.~Nelson}
\email{drnelson@fas.harvard.edu}
\affiliation{Department of Physics, Harvard University, Cambridge, MA 02138, United States of America}

\author{Federico Toschi}
\email{f.toschi@tue.nl}
\affiliation{Department of Applied Physics and Science Education, Eindhoven University of Technology, 5600 MB Eindhoven, The Netherlands}
\affiliation{CNR-IAC, I-00185 Rome, Italy}

\title{The Allee Effect in Compressible Flows}

\begin{abstract}
Microbes in marine environments are often confined to thin near-surface layers while being advected by turbulent flows. Because such constrained advection generates an effectively compressible flow, reproduction and transport interact in a nontrivial way. Here, we focus on populations whose growth is governed by an Allee effect and show that sinks and sources, generated by the compressible flow, have dramatic consequences for the survival of such species.
We derive analytical expressions for the carrying capacity as a function of the Allee strength in the limit of small and large Damk\"ohler number, which measures the product of the large eddy turnover time and the organism growth rate. Numerical simulations reveal how these two limits connect.
In the limit of small Damk\"ohler number, we find a maximal Allee strength, set by the statistics of the compressible flow, that leads to species extinction in fully developed turbulence. 
\end{abstract}

\date{\today}
\maketitle

\textit{Introduction.---}
Microbial populations in oceans and lakes are constantly advected by turbulent flows.
While these tiny organisms reproduce, they are simultaneously being transported, mixed, and dispersed across vast scales.
In large lakes and the open ocean, the large eddy turnover time $\tau_\text{eddy}$ (\textasciitilde$10$ days in the ocean~\cite{Thorpe2005,Vallis2017}) can exceed the typical doubling times of reproducing microbes $\tau_\text{rep}$ (\textasciitilde$12$--$24$ hours for phytoplankton~\cite{Eppley1972}) by more than an order of magnitude.
The latter, however, is still much larger than the turnover time of the smallest eddies, $\propto \sqrt{\eta/\epsilon}$ (\textasciitilde$10$ minutes in the ocean~\cite{Thorpe2005,Vallis2017}), where $\eta$ is the viscosity of water and $\epsilon$ is the energy dissipation rate.
Thus, the biological time scale lies directly within the range over which the turbulent cascade is active, thereby setting the stage for a fascinating interplay between microbial reproduction and turbulent mixing~\cite{dOvidio2010}.

Furthermore, many of these microbes have evolved to localize near the surface (\textasciitilde$10$--$100$ m depth), in thin layers (\textasciitilde$10$ m width), where both light and nutrients are present~\cite{Cullen1982, Cullen2015}.
These layers are much thinner than the large length scales of the turbulent flow, effectively rendering the turbulent dynamics two-dimensional.
However, when floater particles riding on a three-dimensional, incompressible turbulent flow are constrained to a two-dimensional interface, the resulting effective two-dimensional flow field is compressible~\cite{Eckhardt2001, Cressman2003, DePietro2015}.
This effect can be quantified more generally by the dimensionless compressibility
\begin{equation}
\label{eq:kappa}
\kappa=\frac{\langle |\nabla \cdot \boldsymbol{u}|^2 \rangle}{\langle|\nabla \cdot \boldsymbol{u}|^2 \rangle + \langle|\nabla \times \boldsymbol{u}|^2 \rangle},
\end{equation}
where $\boldsymbol{u}$ is the effective two-dimensional flow field, and the brackets $\langle \cdot \rangle$ represent a spatial average.
For passive floaters on top of a turbulently stirred body of water, an effective  compressibility of $\kappa\approx0.5$ was reported~\cite{Cressman2003}, while in simulations of particles constrained to a thin layer within the bulk of isotropic, incompressible, three-dimensional flow one finds $\kappa\approx0.17$~\cite{DePietro2015}.

Microbes in the ocean, however, are not merely passive tracers advected by effectively compressible flows, but are constantly reproducing.
For well-mixed populations,  in an incompressible flow field, the average density $c$ is typically modeled by a first-order differential equation, i.e., $dc/dt=\mu c\, p(c)$, where $\mu=1/\tau_\text{rep}$ and $p(c)$ is the per-capita growth rate.
For simple logistic growth, the per-capita growth decreases linearly with concentration due to competition, i.e., $p(c)=1-c$, where we rescale the population density to set $c=1$ as the carrying capacity of the well-mixed system.
In previous work, the consequences of compressibility for populations with simple logistic growth were studied~\cite{Benzi2009, Perlekar2010, Benzi2022}.
Depending on the Damk\"ohler number $\mathrm{Da}=\tau_\text{eddy}/\tau_\text{rep}$, the carrying capacity can be reduced dramatically, by up to $80\%$ of the saturation concentration.

In this work, we are interested in the effects of compressibility on species for which reproduction is governed not only by competition but also by cooperation.
Indeed, for many species, ranging from mammals and insects to microbes~\cite{Allee1931, Stephens1999, Courchamp2009}, an increase in the per-capita growth rate at low population density (in contrast to the monotonic decrease assumed by the logistic equation) has been reported, often stemming from cooperative effects between individuals.
This phenomenon is known as the Allee effect and is modeled, in a minimal manner, by a quadratic form for the per-capita growth rate
\begin{equation}
    p(c)=(1-c)(c-a),
\end{equation}
where $a$ is the Allee strength.
Depending on $a$, there are four interesting regimes:
(i) $a<-1$: $p(c)$ is monotonically decreasing in the biologically relevant domain $0<c<1$, similar to logistic growth where $p(c)=1-c$;
(ii) $-1<a<0$: $p(c)>0$ and has positive slope for $c\gtrsim0$, known as a weak Allee effect;
(iii) $0<a<1$: $p(c)<0$ but has positive slope for $c\gtrsim0$, known as a strong Allee effect;
(iv) $a>1$: $p(c)<0$ in the relevant domain $0<c<1$, so that the population always goes extinct.

Crucially, for a strong Allee effect, a critical population density $c^*=a$ is necessary for growth, and below which, the well-mixed population goes extinct. In yeast cells (\emph{S. cerevisiae}), for example, such a strong Allee effect was found due to a cooperative sucrose metabolism~\cite{Gore2009}.
More relevant in the context of this study, a strong Allee effect has also been reported for \emph{V. fischeri}~\cite{Kaul2016}, a widespread marine bacterium. Even without predation, an Allee effect was found, potentially stemming from a quorum sensing mechanism such that cells need to reach a certain density to switch on growth‑promoting functions. The critical population density became higher, i.e., the Allee effect became stronger, in the presence of predation due to a bacterivore (\emph{Cafeteria roenbergensis}), where at low densities, the individual risk of predation of a single bacterium is much bigger than at higher densities.
Furthermore, in recent years, quorum sensing has been found in many marine bacteria~\cite{Coolahan2025}. For example, sexual reproduction occurs during the natural bloom of a widespread, toxic single-celled alga (\emph{Pseudo‑nitzschia australis})~\cite{Prigent2025}. Both mechanisms provide motivation for inducing an Allee effect in these marine microbes.

\textit{Dynamics.---}
We approximate microbes with an Allee effect confined to the upper layers of the ocean by a two-dimensional convection--diffusion equation in a compressible flow with local reproduction,
\begin{equation}
    \label{eq:dcdt}
    \partial_t c + \boldsymbol{\nabla}\cdot(\boldsymbol{u} c) = D \nabla^2 c + \mu c (1-c)(c-a),
\end{equation}
where $\boldsymbol{u}(\boldsymbol{x},t)$ is a two-dimensional compressible flow field and $D$ is the diffusivity of the microbes.
Reproduction follows the standard local form of the Allee effect, with Allee strength $a$ and reproduction rate $\mu = 1/\tau_\text{rep}$.

Note that, in the absence of advection and for a strong Allee effect, i.e., $0<a<1$, this dynamics reduces to a deterministic version of model A in the classification of critical phenomena~\cite{Hohenberg1977}, also known as the Allen--Cahn equation~\cite{Allen1979, Evans1992}.
This model describes the relaxation of a non-conserved order parameter and can be written as $\partial_t c=-D\,\delta F/\delta c$, with the free energy 
\begin{gather}
\label{eq:F}
    F=\int d\mathbf{x}\,\left[f(c)+|\nabla c|^2\right], \\
    f(c)=\mu c^2 \big(3c^2-4(1+a)c+6a\big)/(12D),
    \label{eq:f_den}
\end{gather}
where $f$ is the local free energy density.
The latter function is a double-well potential with minima at $c=0$ and $c=1$.
For $a=1/2$, both minima have the same free energy, while for $a<1/2$ ($a>1/2$) the state $c=1$ ($c=0$) is the true equilibrium state.
In the following, we show that compressible turbulent flows significantly alter this stability.

\begin{figure}[!t]
\includegraphics[width=0.8\linewidth]{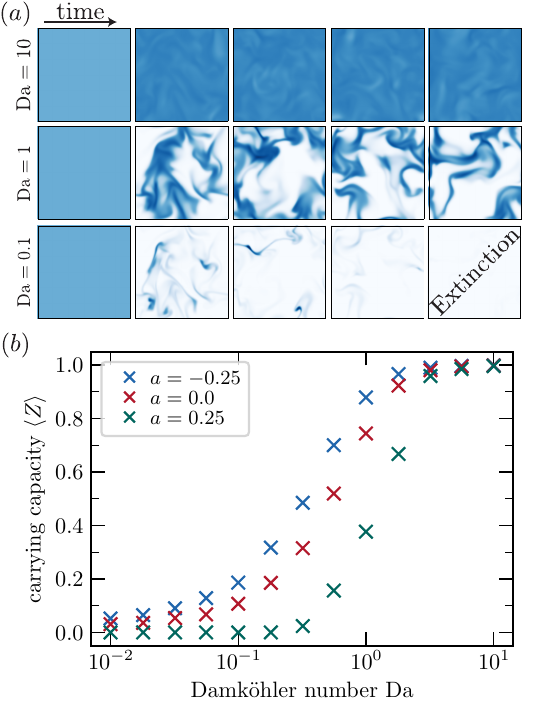}
    \caption{
    \textbf{Dynamics across Damk\"ohler numbers.}
    (a) Concentration fields at different times for representative simulations with $a=0.25$ and $\mathrm{Da}=0.1,1,10$.
    (b) Long-time averaged stationary carrying capacity $\langle Z\rangle$ as a function of the Damk\"ohler number for three different values of the Allee strength $a$.
    Note the dramatic decrease at small carrying capacity with an extinction transition when $a=0.25$ and $\kappa \approx 0.17$.
    For numerical details, see the Supplemental Materials.
    }
    \label{fig:fig1}
\end{figure}

To study compressible flow numerically, we solve the Navier--Stokes equations in a three-dimensional, incompressible setting with random large-scale forcing with periodic boundaries, such that the flow is isotropic.
From this three-dimensional flow field, we follow Ref.~\cite{Perlekar2010} and select one layer at a mid-plane of the periodic cube and construct a projected two-dimensional flow field $\boldsymbol{u}(\boldsymbol{x},t)=(u_x(\boldsymbol{x},t),u_y(\boldsymbol{x},t))$, at position $\boldsymbol{x}=(x,y)$ and time $t$, such that $\boldsymbol{u}(\boldsymbol{x},t)$ is compressible with $\kappa\approx0.17$.
In Fig.~\ref{fig:fig1}(a), we show concentration fields at different times (in columns) for representative simulations with Damk\"ohler numbers $\mathrm{Da}=0.1,1,10$ (in rows), with periodic boundary conditions in $x$ and $y$, and for a strong Allee effect, $a=0.25$.
Initially (first column), we set $c(\mathbf{x},t=0)=0.75$ homogeneously.

For $\mathrm{Da}=10$, corresponding to $\tau_\text{eddy}\gg\tau_\text{rep}$, reproduction is fast enough that the population can respond to the compressible flow almost instantaneously.
Thus, the concentration remains almost everywhere close to the carrying capacity of the well-mixed system, which we use as the reference concentration, i.e., $c(\mathbf{x},t)\approx1$ for all $\mathbf{x}$ and $t$, with only small fluctuations in space induced by the slowly varying compressible flow.
Consequently, the averaged carrying capacity of the spatial system, defined as
\begin{equation}
\label{eq:carry}
    Z(t)=\frac{1}{L^2}\int d\mathbf{x}\; c(\mathbf{x},t),
\end{equation}
where $L$ is the system size, also remains close to that of the well-mixed system, i.e., $\langle Z(t)\rangle\approx1$.
This is also seen in Fig.~\ref{fig:fig1}(b), where we plot the averaged carrying capacity for three different values of $a$ as a function of the Damk\"ohler number.

When the Damk\"ohler number approaches unity, the dynamics changes significantly.
Here, the eddy turnover time and the reproduction time are of the same order, and there is a persistent interplay between local compression into sinks and transport away from sources of the flow.
We observe striking filamentary spatial patterns similar to those known for passive floaters~\cite{Cressman2003, Qi2025}.
As for populations undergoing logistic growth, the averaged carrying capacity is strongly reduced in this regime~\cite{Benzi2009, Perlekar2010}.
However, for the present Allee dynamics, this reduction becomes more pronounced as the Allee effect becomes stronger, as shown in Fig.~\ref{fig:fig1}(b).
The carrying capacities shown there are averaged over long times and correspond to statistically stationary states with only small fluctuations.

Finally, for sufficiently small Damk\"ohler number $\mathrm{Da}=\mu \tau_\text{eddy}$, a strong Allee effect can lead to extinction.
This is illustrated by the last column of Fig.~\ref{fig:fig1}(a) for $\mathrm{Da}=0.1$, where the effect of the compressible flow is strong enough to eventually drive the population to collapse.
Importantly, the same collapse also occurs when the system is initialized everywhere at the full carrying capacity of the well-mixed system.
Thus, extinction is not merely a consequence of unfavorable initial conditions, but is dynamically induced by the interplay of slow reproduction and compressible transport.

\begin{figure}[!t]
    \includegraphics[width=0.85\linewidth]{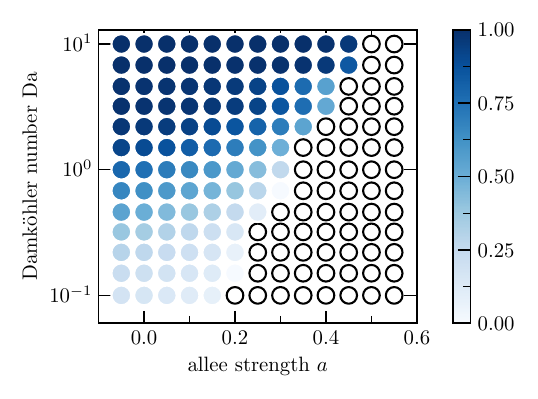}
    \caption{
    \textbf{Stationary carrying capacity in the $(a,\mathrm{Da})$-plane.}
    Long-time averaged stationary carrying capacity $\langle Z\rangle$ as a function of Allee strength $a$ and Damk\"ohler number $\mathrm{Da}$.
    Open black circles indicate extinction, while the blue scale denotes the stationary carrying capacity in surviving states.
    For numerical details, see the Supplementary Materials.
    }
    \label{fig:fig2}
\end{figure}

Figure~\ref{fig:fig2} summarizes the long time, stationary behavior as the Allee strength and Damk\"ohler number are varied in the $(a,\mathrm{Da})$-plane for $\kappa\approx 0.17$.
For small or negative Allee strength, the population survives for all Damk\"ohler numbers, although the carrying capacity is strongly suppressed at intermediate and small $\mathrm{Da}$.
As $a$ increases, the carrying capacity is reduced until an extinction boundary emerges.
This boundary, a kind of first-order phase transition, separates a region of nonzero stationary carrying capacity from a region in which the population always goes extinct, i.e., $\langle Z \rangle = 0$.
Two trends are clearly visible.
First, for fixed $a$, survival becomes more likely as $\mathrm{Da} =\mu \tau_{eddy}$ increases, because faster local reproduction allows the population to recover from compressive depletion near transient sinks before it is swept below the local threshold.
Second, for fixed $\mathrm{Da}$, survival becomes less likely as $a$ increases, because the critical density required for local growth becomes harder to maintain in a strongly inhomogeneous concentration field.
The phase boundary therefore interpolates between two limiting mechanisms: at large $\mathrm{Da}$ it approaches the equilibrium threshold at $a=1/2$, while at small $\mathrm{Da}$ it is governed by the statistics of passive clustering in the compressible flow.

A full analytical treatment of this mixing problem for a non-conserved scalar in turbulent compressible flow is a challenging problem.
However, progress can be made in the two limiting cases of large and small Damk\"ohler number.
In the following, we discuss these two limits before returning to the role of compressibility itself.

\textit{Large Damk\"ohler number limit.---}
When $\tau_\text{eddy}\gg\tau_\text{rep}$, reproduction can compensate for any inhomogeneities caused by local sinks and sources such that the concentration remains locally close to the stationary fixed point at $c=1$. In this limit, we can perturbatively expand the concentration field in powers of $\tau_\text{rep} = 1/\mu$ as
\begin{equation}
    c(\mathbf{x},t)=\sum_{n=0}^{\infty}\tau_\text{rep}^n c_n(\mathbf{x},t),
    \label{eq:exp}
\end{equation}
with $c_0=1$.
Using this expansion in Eq.~\eqref{eq:dcdt} and sorting terms order by order in $\tau_\text{rep}$, we obtain for the averaged stationary carrying capacity, to leading order,
\begin{equation}
\label{eq:cor_z_large}
    \langle Z\rangle = 1 - \frac{\tau_\text{rep}^2}{L^2}\frac{2-a}{(1-a)^3}
    \Big\langle \int d \mathbf{x} \; \big(\nabla \cdot \mathbf{u}\big)^2 \Big\rangle
    + \mathcal{O}(\tau_\text{rep}^3).
\end{equation}
Thus, the carrying capacity is only weakly reduced from the otherwise stable state $c=1$, when the reproduction rate $\mu = 1/\tau_\text{rep}$ is large, and this reduction becomes stronger as $a$ grows.

This perspective, however, is only valid for $a<1/2$.
As mentioned above, in this regime the state $c=1$ is the more stable fixed point, and even if the compressible flow manages to deplete the concentration locally so that a domain with $c\approx0$ emerges, a pushed wave tends to close such domains~\cite{Benzi2022}.
For $a>1/2$, however, the state $c=0$ becomes more stable than $c=1$, and whenever compressibility manages to create a sufficiently large depleted region (thus overcoming a genetic line tension~\cite{Benzi2022}), this region expands rather than closes.
In that case, fluctuations induced by the compressible flow can nucleate empty domains that subsequently grow and eventually drive the entire population to extinction, similar to front propagation under mixing~\cite{Mahoney2012, Clavin2016}.
The large-Damk\"ohler regime is therefore controlled by two effects: weak perturbative suppression of the carrying capacity for $a\lesssim 1/2$, and nucleation-assisted collapse for $a\gtrsim 1/2$.

\textit{Small Damk\"ohler limit.---}
In the opposite limit, $\tau_\text{rep}\gg\tau_\text{eddy}$, the concentration field is strongly inhomogeneous in space, as is evident from the bottom panel of Fig.~\ref{fig:fig1}(a).
By taking the temporal derivative of Eq.~\eqref{eq:carry} and using Eq.~\eqref{eq:dcdt}, we find that the carrying capacity can be constant in time only when
\begin{equation}
\label{eq:stat_c}
    \langle c (1-c)(c-a)\rangle = 0.
\end{equation}
Thus, on average, growth and decay must balance within the system.

In the limit $\tau_\text{rep}\to\infty$, the concentration of individuals behaves approximately as the density of a set of passive Lagrangian tracer particles advected by a compressible flow, up to an overall scale set by the carrying capacity.
Following Ref.~\cite{Perlekar2010}, we therefore write
\begin{equation}
\label{eq:ansatz_c}
    c(\mathbf{x},t)\approx L^2 \langle Z \rangle \mathcal{P}(\mathbf{x},t),
\end{equation}
where $\mathcal{P}(\mathbf{x},t)$ is the probability density of finding such a particle at position $\mathbf{x}$ and time $t$.
For passive particles, the density $\mathcal{P}$ evolves according to
\begin{equation}
\label{eq:fkkp}
    \partial_t \mathcal{P} + \nabla \cdot ( \mathbf{u}  \mathcal{P}) = D \nabla^2 \mathcal{P}.
\end{equation}
This is the standard advection--diffusion problem for passive floaters in compressible flow.
Due to compressibility, multifractal clustering occurs and $\mathcal{P}$ is not spatially uniform~\cite{Balkovsky2001, Falkovich2001}.
This can be seen by multiplying Eq.~\eqref{eq:fkkp} by $n\mathcal{P}^{n-1}$ and averaging in space to obtain
\begin{align}
\label{eq:moments_dyn}
    \frac{d}{dt}\int \mathcal{P}^n d\mathbf{x}
    &=
-(n-1)\int (\nabla\cdot \mathbf{u})\,\mathcal{P}^n d\mathbf{x}
\nonumber\\
&\hspace{1cm}
- D n(n-1)\int \mathcal{P}^{\,n-2}\,|\nabla \mathcal{P}|^2 d\mathbf{x}.
\end{align}
For $n=1$, $\int \mathcal{P}\, d\mathbf{x}=1$ by conservation of probability.
For $n>1$, however, the diffusive term is strictly negative and tends to suppress fluctuations, whereas local sinks with $(\nabla\cdot\mathbf{u})<0$ that coincide with regions of large $\mathcal{P}$ increase higher moments.
In a statistically stationary state, such that the left side of Eq.~\eqref{eq:moments_dyn} vanishes,  both contributions balance.
We denote the stationary moments associated with Eq.~\eqref{eq:moments_dyn} by $\langle \mathcal{P}^n\rangle$.

Upon substituting Eq.~\eqref{eq:ansatz_c} in Eq.~\eqref{eq:stat_c}, we find that in the low-Damk\"ohler regime the stationary carrying capacity must be given by either $\langle Z\rangle=0$ or
\begin{gather}
\label{eq:car_lowDa}
    \langle Z\rangle^\pm = \frac{\langle \mathcal{P}\rangle \langle \mathcal{P}^2\rangle}{2 \langle \mathcal{P}^3\rangle}
    \left[
    (1+a) \pm \sqrt{(1+a)^2 - 4 a \gamma}
    \right], \\
    \text{with } \gamma \equiv \frac{\langle \mathcal{P}\rangle \langle \mathcal{P}^3\rangle}{\langle \mathcal{P}^2\rangle^2}.
\end{gather}
Thus, the carrying capacity depends in a nontrivial way on the first three moments of $\mathcal{P}$ and on the Allee strength $a$.
We note the Cauchy--Schwarz inequality ensures
\begin{equation}
\gamma\geq1,
\end{equation}
with equality holding only when $\mathcal{P}$ is spatially uniform~\footnote{As can be seen using $|\langle XY\rangle |^2\leq \langle X^2\rangle \langle Y^2\rangle$ with $X=\mathcal{P}^{1/2}$ and $Y=\mathcal{P}^{3/2}$, and $\mathcal{P}>0$, everywhere.}.
In the incompressible case, where $\mathcal{P}$ is constant in space and $\langle \mathcal{P}^n\rangle=\langle \mathcal{P}\rangle^n$, one recovers $\langle Z\rangle^-=a$ and $\langle Z\rangle^+=1$, which coincide with the unstable and one of the locally stable fixed points of the well-mixed system (the other locally stable fixed point corresponds to $\langle Z \rangle =0$).
Compressibility, however, leads to spatial fluctuations of $\mathcal{P}$ and hence importantly shifts the regimes where $\langle Z \rangle = 0$ and $\langle Z \rangle$ is nonzero.

\begin{figure}[!t]
    \includegraphics[width=0.8\linewidth]{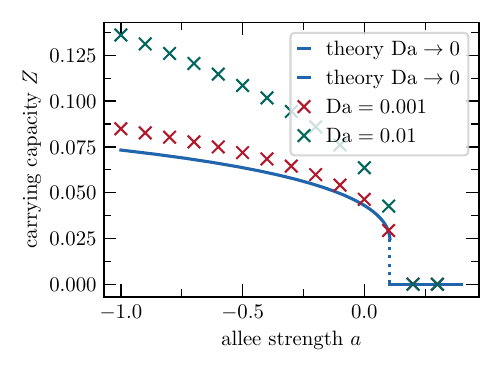}
    \caption{
    \textbf{Low-Damk\"ohler theory and simulations.}
    Stationary carrying capacity $\langle Z\rangle$ as a function of the Allee strength $a$ in the low-Damk\"ohler regime.
    The solid line shows the theoretical prediction for the carrying capacity $\langle Z\rangle^{+}$ from Eq.~\eqref{eq:car_lowDa}, while symbols show numerical results for small but finite Damk\"ohler numbers.
    For numerical details, see the Supplemental Materials.
    }
    \label{fig:fig3}
\end{figure}

For the square root in Eq.~\eqref{eq:car_lowDa} to be real and represent a real carrying capacity, we require
\begin{equation}
    \frac{(1+a)^2}{4a}\geq \gamma,
\end{equation}
which yields a critical value of the Allee strength,
\begin{equation}
\label{eq:acrit}
    a_\text{crit}^{\pm}=
    2 \gamma -1 \pm 2 \sqrt{\gamma (\gamma -1)}
\end{equation}
For a spatially homogeneous field, with $\gamma = 1$, we find $a_\text{crit}^{\pm}=1$, i.e., the upper bound of the Allee strength for well-mixed systems. 
However, when $\mathcal{P}$ fluctuates in space, $\gamma >1$ and $a_\text{crit}^{+}>1$ and $0<a_\text{crit}^{-}<1$.
In the biologically relevant strong-Allee regime $0<a<1$, the lower solution $a_\text{crit}^{-}$ sets the relevant extinction threshold.
In the limit of strong spatial fluctuations, where
$\gamma \to\infty$, one finds $a_\text{crit}^{-}\approx 1/(4\gamma)\to0$, demonstrating that strong compressibility can exclude populations even for very weak Allee effects.
Furthermore, note that there is a jump in the carrying capacity at the critical Allee strength $a_\text{crit}^{-}$, i.e.,
\begin{equation}
    \langle Z \rangle^\pm|_{a=a_\text{crit}^{-}} = \frac{\langle \mathcal{P}\rangle^2}{\langle \mathcal{P}^2\rangle} \left( 1 -\sqrt{1-\gamma^{-1}} \right)
\end{equation}
reminiscent of a first order phase transition, that survives even when the Damk\"ohler number $\text{Da}=\mu \tau_\text{eddy}$ is small, i.e., in the limit of strong compressible turbulence.

Upon using the compressible flow field constructed in the same manner as for Fig.~\ref{fig:fig1}(a) in Eq.~\eqref{eq:fkkp}, we numerically determine $\langle \mathcal{P}^2\rangle/\langle \mathcal{P}\rangle^2=8$ and $\langle \mathcal{P}^3\rangle/\langle \mathcal{P}\rangle^3=186$, where we use the same Schmidt number $\mathrm{Sc}=\nu/D$ as in the simulations in the low-Damk\"ohler regime.
This yields a critical value $a_\text{crit}^{-}\approx0.1$ with the carrying capacity $\langle Z \rangle^\pm|_{a=a_\text{crit}^{-}} = 0.025$.
In Fig.~\ref{fig:fig3}, we compare the positive branch $\langle Z\rangle^{+}$ of Eq.~\eqref{eq:car_lowDa} with numerically measured carrying capacities for small Damk\"ohler numbers.
As $\mathrm{Da}\to0$, the numerical data support the theoretical prediction given by Eq.~\eqref{eq:car_lowDa} for the average carrying capacity $\langle Z \rangle $.
Thus, in the slow-reproduction regime, the extinction threshold is indeed set by the clustering statistics generated by the compressible flow.
While the clustering reduces the carrying capacity, remarkably, it also allows for the invasion of a population even for homogeneous initial conditions far below the critical Allee strength of the well-mixed systems, see the Supplementary Materials.

\textit{Changing the compressibility.---}
So far, we have kept the flow statistics fixed and varied only the ratio of transport and reproduction time scales and the Allee strength.
A complementary question is how the phase behavior changes when the degree of compressibility itself is varied.
By decomposing the turbulent flow field into an incompressible $\mathbf{u}_i$ and compressible part $\mathbf{u}_c$, we define a continuous family of flow fields
\begin{equation}
    \mathbf{\tilde{u}}(\mathbf{x},t) = \sqrt{2}\left[ \mathbf{u}_i \cos(\pi \beta/2) + \mathbf{u}_c \sin(\pi \beta/2) \right],
\end{equation}
where $\beta \in [0,1]$ is a compressibility parameter~\cite{Boffetta2004, Perlekar2013}. The latter allows for a continuous variation of the compressibility $\kappa$, while leaving the basic turbulent structure of the flow unchanged. 
The resulting compressibility $\kappa(\beta)$, defined by Eq.~\eqref{eq:kappa}, increases monotonically with $\beta$ from $\kappa=0$ for $\beta=0$ to $\kappa=1$ for $\beta=1$~\cite{Perlekar2013}.

\begin{figure}[!t]
\includegraphics[width=0.8\linewidth]{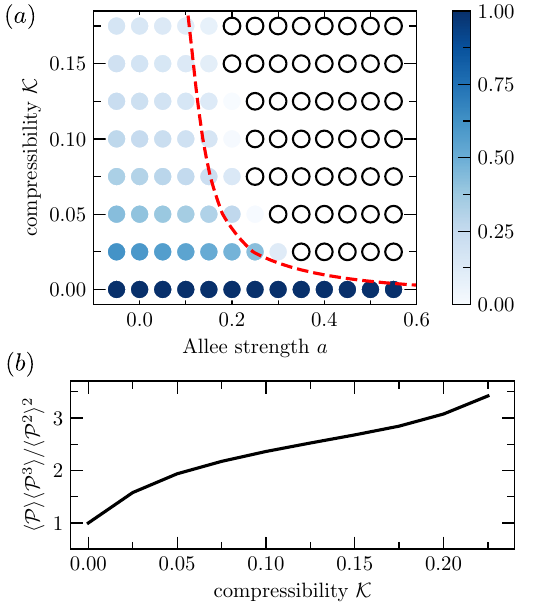}
    \caption{
    \textbf{Effects of varying the compressibility.}
    (a) Stationary carrying capacity in the plane spanned by Allee strength $a$ and compressibility $\kappa(\beta)$, for fixed low Damk\"ohler number $\text{Da}=0.1$.
    Black circles indicate extinction; the dashed red line marks the critical Allee strength $a_c^-(\kappa)$.
    (b) Effective compressibility $\kappa$ as a function of $\beta$.
    For numerical details, see the Supplemental Materials.
    }
    \label{fig:fig4}
\end{figure}

Using this construction, we fix $\beta$ such that the flow corresponds to a compressibility and show the stationary carrying capacity in the $(a,\kappa)$-plane, for a fixed Damk\"ohler number $\text{Da}=0.1$, in Fig.~\ref{fig:fig4}(a). We find that increasing the compressibility $\kappa$ systematically suppresses survival.
For an incompressible flow ($\kappa = 0$), the system can be understood as well-mixed and even for Allee strengths up to $a=1$ populations can survive as long as the initial conditions are larger than this $a$. The dependence on the initial condition, i.e., on how $c$ is initially distributed can be understood using an analysis similar to that used for the fixation of antagonistic species with a selective advantage in incompressible flow~\cite{Bauermann2025}.
However, even for weakly compressible flow, the survival region shrinks dramatically and the extinction boundary moves toward smaller values of $a$.
This is precisely what one expects from the low-Damk\"ohler number theory: compressibility enhances clustering of passive trajectories, increases the higher moments of $\mathcal{P}$, and thereby decreases the relevant critical value $a_\text{crit}^{-}$. 
In Fig.~\ref{fig:fig4}(b), we show the measured ratios of $\gamma = \langle \mathcal{P}\rangle \langle \mathcal{P}^3\rangle/\langle \mathcal{P}^2\rangle^2$, which monotonically increases as a function of $\kappa$. We can now use these ratios in Eq.~\eqref{eq:acrit} for predicting the maximal Allee strength $a_\text{crit}^{-}$ in the low-Damk\"ohler regime (red, dashed line in Fig.~\ref{fig:fig4}(a)). 
This figure demonstrates that even when compressibility modestly increases the ratio $\langle \mathcal{P}\rangle \langle \mathcal{P}^3\rangle/\langle \mathcal{P}^2\rangle^2$, the critical Allee strength is drastically reduced.
Below the critical line in Fig.~\ref{fig:fig4}(a), the system settles into an inhomogeneous but statistical steady state with nonzero carrying capacity.
Above it, local depletion events become sufficiently strong and sufficiently frequent that the system is driven irreversibly toward extinction.
Thus, the compressibility also acts as a control parameter for the loss of population stability.

\textit{Discussion.---}
Our results show that compressible turbulent transport can qualitatively change the stability of populations subject to an Allee effect.
For simple logistic growth, compressibility simply reduces the carrying capacity.
However, for bistable growth with a strong Allee effect the compressibility can become fatal.

In the large-Damk\"ohler number regime, reproduction is fast enough that the concentration field remains close to the stable well-mixed state $c=1$.
Here, compressibility produces only the perturbative correction shown in Eq.~\eqref{eq:cor_z_large}, and the stationary carrying capacity remains close to unity.
Nevertheless, this correction grows with the Allee strength, showing that populations with an Allee effect are already more vulnerable than logistic ones even when growth is fast.
For $a>1/2$, the character of the dynamics changes because the empty state becomes favorable, as shown in Eqs.~\eqref{eq:F} and ~\eqref{eq:f_den}.
In that case, compressibility can nucleate sufficiently large depleted domains that subsequently grow as pushed waves and destroy the population.

In the opposite, small-Damk\"ohler regime, the population fate is controlled by the passive clustering statistics of the compressible flow.
The concentration field becomes slaved to the density of advected particles, and the carrying capacity is determined by the first three moments of their probability measure.
This observation yields an explicit prediction for the surviving branch and, more importantly, a critical Allee strength $a_\text{crit}^{-}$ necessary for the population to persist.
Furthermore, we find that this critical Allee strength is strongly reduced even for slightly compressible flows.

More generally, our results suggest that for populations with cooperative growth, turbulent transport cannot be understood as a mere renormalization of mixing or diffusivity. 
Instead, compressible stirring changes the effective stability landscape of the population, and sets an upper limit on the Allee strength for marine bacteria that exhibit this phenomenon~\cite{Kaul2016}.
It seems plausible that similar phenomena occur not only for microbial communities in aquatic environments, but also more generally for populations constrained to thin layers, interfaces, or active transport networks, where compressibility and local cooperation coexist.
An important direction for future work will be to include demographic noise, which may further enhance extinction in the vicinity of the thresholds identified here.

\section*{Acknowledgment}
The authors thank Tali Khain for fruitful discussions.

\newpage

\section*{Supplementary Materials}
\subsection{Numerical details}
We generate the turbulent velocity field by first solving the classical incompressible three-dimensional Navier--Stokes equation with random forcing at large scales, using a pseudospectral method.
This gives $\vec{\mathbf{u}}(\vec{\mathbf{r}}, t) = (u_x(\vec{\mathbf{r}},t), u_y(\vec{\mathbf{r}},t), u_z(\vec{\mathbf{r}},t))$ at position $\vec{\mathbf{r}}=(x, y, z)$ and time $t$.
In all simulations, we use the same resolution ($N = 128$ grid points for a system size $L=2 \pi$) and periodic boundary conditions as in the numerical solver for the advected reaction--diffusion equations.
For the compressible advection velocity in Eq.~\eqref{eq:dcdt}, we use $\vec{\mathbf{u}}^{2d}(x, y, t) = (u_x (x,y, z=L/2,t), u_y (x,y, z=L/2,t))$ as the projection  midplane of the periodic cube correspondingthe the 3d simulation. This two-dimensional, compressible flow field is used to update the concentration field $c(x,y,t)$ with a finite-difference solver using an Adams--Bashforth time-integration scheme.

\subsection{Compressibility can help populations invade from homogeneous initial conditions}
In homogeneous systems (described by Eq.~\eqref{eq:dcdt} with all gradient terms set to zero) initialized below the critical Allee strength, the population relaxes exponentially toward the stable fixed point at $c=0$. However, at low Damk\"ohler numbers, compressible flow generates concentrated filamentary spatial patterns, and the concentration near local sinks can exceed the critical Allee level. Thus, even for homogeneous initial concentrations below the threshold, compressibility can allow the population to survive.
\begin{figure}[!t]
\includegraphics[width=0.8\linewidth]{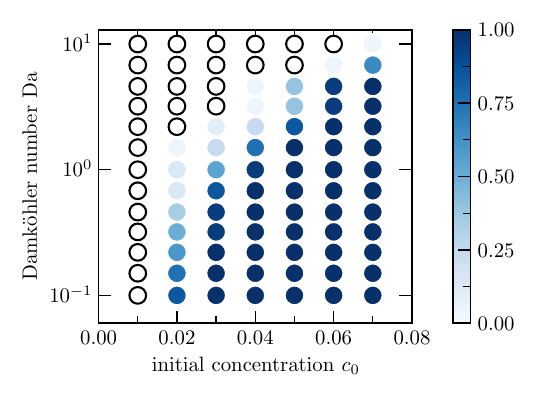}
    \caption{
    \textbf{Invasion probability for homogeneous initial conditions} below the critical Allee strength, averaged over 20 independent runs with $a=0.1$.
    Extinction is marked by open black circles.
    }
    \label{fig:fig_sm}
\end{figure}
This remarkable effect of compressible turbulence is illustrated in Fig.~\ref{fig:fig_sm}, where we plot the probability for invasion over $20$ independent runs for each combination of Damk\"ohler number and homogeneous initial concentration $c_0<a$, with $a=0.1$.
For low Damk\"ohler numbers, where filamentary spatial patterns occur, the population is able to invade the system in some realizations even for initial conditions as low as $c(\mathbf{x},t=0)=0.02$, and then settles down at its reduced carrying capacity.
For large Damk\"ohler numbers, by contrast, compressibility does not induce sufficiently strong spatial inhomogeneities.
Thus, in this limit, the population can invade only when the initial concentration is above the critical Allee strength $c=a$.

\end{document}